# Design and performance of wide-band corrugated walls for the BICEP Array detector modules at 30/40 GHz

A. Soliman[a], P. A. R. Ade[b], Z. Ahmed[c], R. W. Aikin[a], K. D. Alexander[d], D. Barkats[d], S. J. Benton[e], C. A. Bischoff[f], J. J. Bock[a,g], R. Bowens-Rubin[d], J. A. Brevik[a], I. Buder[d], E. Bullock[h], V. Buza[d,i], J. Connors[d], J. Cornelison[d], B. P. Crill[g], M. Crumrine[j], M. Dierickx[d], L. Duband[w], C. Dvorkin[i], J. P. Filippini[l,m], S. Fliescher[j], J. Grayson[k], G. Hall[j], M. Halpern[n], S. Harrison[d], S. R. Hildebrandt[a,g], G. C. Hilton[o], H. Hui[a], K. D. Irwin[k,c,o], J. H. Kang[k], K. S. Karkare[d,r], E. Karpel[k], J. P. Kaufman[q], B. G. Keating[q], S. Kefeli[a], S. A. Kernasovskiy[k], J. M. Kovac[d,i], C. L. Kuo[k,c], K. Lau[j], N. A. Larsen[r], E. M. Leitch[r], M. Lueker[a], K. G. Megerian[g], L. Moncelsi[a], T. Namikawa[s], C. B. Netterfield[e,t], H. T. Nguyen[g], R. O'Brient[a,g], R. W. Ogburn IV[k,c], S. Palladino[f], C. Pryke[h,j], B. Racine[d], S. Richter[d], R. Schwarz[h], A. Schillaci[a], C. D. Sheehy[u], T. St. Germaine[d], Z. K. Staniszewski[a,g], B. Steinbach[a], R. V. Sudiwala[b], G. P. Teply[a,q], K. L. Thompson[k,c], J. E. Tolan[k], C. Tucker[b], A. D. Turner[g], C. Umilt`a[f], J. Van Zyl[a], A. G. Vieregg[v,r], A. Wandui[a], A. C. Weber[g], D. V. Wiebe[b], J. Willmert[h], C. L. Wong[d,i], W. L. K. Wu[r], E. Yang[c], K. W. Yoon[c,k], and C. Zhang[a]

[a]Department of Physics, California Institute of Technology, Pasadena, California 91125, USA
[b]School of Physics and Astronomy, Cardiff University, Cardiff, CF24 3AA, United Kingdom
[c]Kavli Institute for Particle Astrophysics and Cosmology, SLAC National Accelerator Laboratory, 2575 Sand Hill Rd, Menlo Park, California 94025, USA
[d]Harvard-Smithsonian Center for Astrophysics, 60 Garden Street MS 42, Cambridge, Massachusetts 02138, USA
[e]Department of Physics, University of Toronto, Toronto, Ontario, M5S 1A7, Canada
[f]Department of Physics, University of Cincinnati, Cincinnati, Ohio 45221, USA
[g]Jet Propulsion Laboratory, Pasadena, California 91109, USA
[h]Minnesota Institute for Astrophysics, University of Minnesota, Minneapolis, Minnesota 55455, USA
[i]Department of Physics, Harvard University, Cambridge, MA 02138, USA
[j]School of Physics and Astronomy, University of Minnesota, Minneapolis, Minnesota 55455, USA
[w]Service des Basses Températures, Commissariat `a l'Energie Atomique, 38054 Grenoble, France
[l]Department of Physics, University of Illinois at Urbana-Champaign, Urbana, Illinois 61801, USA
[m]Department of Astronomy, University of Illinois at Urbana-Champaign, Urbana, Illinois 61801, USA
[k]Department of Physics, Stanford University, Stanford, California 94305, USA
[n]Department of Physics and Astronomy, University of British Columbia, Vancouver, British Columbia, V6T 1Z1, Canada
[o]National Institute of Standards and Technology, Boulder, Colorado 80305, USA
[q]Department of Physics, University of California at San Diego, La Jolla, California 92093, USA
[r]Kavli Institute of Cosmological Physics, University of Chicago, Chicago, IL 60637, USA
[s]Leung Center for Cosmology and Particle Astrophysics, National Taiwan University, Taipei 10617, Taiwan
[t]Canadian Institute for Advanced Research, Toronto, Ontario, M5G 1Z8, Canada
[u]Physics Department, Brookhaven National Laboratory, Upton, NY 11973
[v]Department of Physics, Enrico Fermi Institute, University of Chicago, Chicago, IL 60637, USA

**ABSTRACT**

BICEP Array is a degree-scale Cosmic Microwave Background (CMB) experiment that will search for primordial B-mode polarization while constraining Galactic foregrounds. BICEP Array will be comprised of four receivers to cover a broad frequency range with channels at 30/40, 95, 150 and 220/270 GHz. The first low-frequency receiver will map synchrotron emission at 30 and 40 GHz and will deploy to the South Pole at the end of 2019. In this paper, we give an overview of the BICEP Array science and instrument, with a focus on the detector module. We designed corrugations in the metal frame of the module to suppress unwanted interactions with the antenna-coupled detectors that would otherwise deform the beams of edge pixels. This design reduces the residual beam systematics and temperature-to-polarization leakage due to beam steering and shape mismatch between polarized beam pairs. We report on the simulated performance of single- and wide-band corrugations designed to minimize these effects. Our optimized design alleviates beam differential ellipticity caused by the metal frame to about 7% over 57% bandwidth (25 to 45 GHz), which is close to the level due the bare antenna itself without a metal frame. Initial laboratory measurements are also presented.

## 1. INTRODUCTION

Many inflation models predict a primordial gravitational-wave background that imprints B-mode polarization in the Cosmic Microwave Background (CMB). The amplitude of the inflationary B-modes depends on the energy scale of inflation and is parametrized by the tensor-to-scalar ratio ($r$). High sensitivity instruments are necessary to constrain the B-mode amplitude with wide frequency coverage to remove Galactic foreground. The current BICEP/Keck experiments [1-4] constraint inflationary gravitational waves at $r_{0.05}$ < 0.07 at 95% confidence, when combined with constraints from Planck CMB temperature plus baryon acoustic oscillations.

BICEP Array is the next CMB experiment in the BICEP/Keck series, providing high-sensitivity with channels at 30/40, 95, 150 and 220/270 GHz. The highest and lowest frequency receivers operate in two bands (30/40 and 220/270 GHz) simultaneously, using two detector designs over the focal plane. The BICEP Array will measure the polarized sky in different frequency bands to reach an ultimate sensitivity to the amplitude of gravitational waves of 0.002 < $\sigma(r)$ < 0.006, depending on the level of removal of the B-mode gravitational lensing signal, and after subtracting the polarized Galactic synchrotron and dust emission. Channels at low frequency are necessary to measure and model Galactic synchrotron emission, which we project will be the limiting foreground once all the higher-frequency data already in our possession will be analyzed. The low-frequency BICEP Array receiver will be a new design operating in two distinct bands centered at 30 and 40 GHz, which will be deployed to the South Pole at the end of 2019. The BICEP Array receivers are based on the highly successful BICEP3 and Keck Array receiver designs [3], and feature a 550mm-aperture refractive telescope, pulse-tube cooled to 4K inside a dedicated cryostat, as shown in Figure 1. Details of the receivers and mount are given in the concurrently presented references [5] and [6].

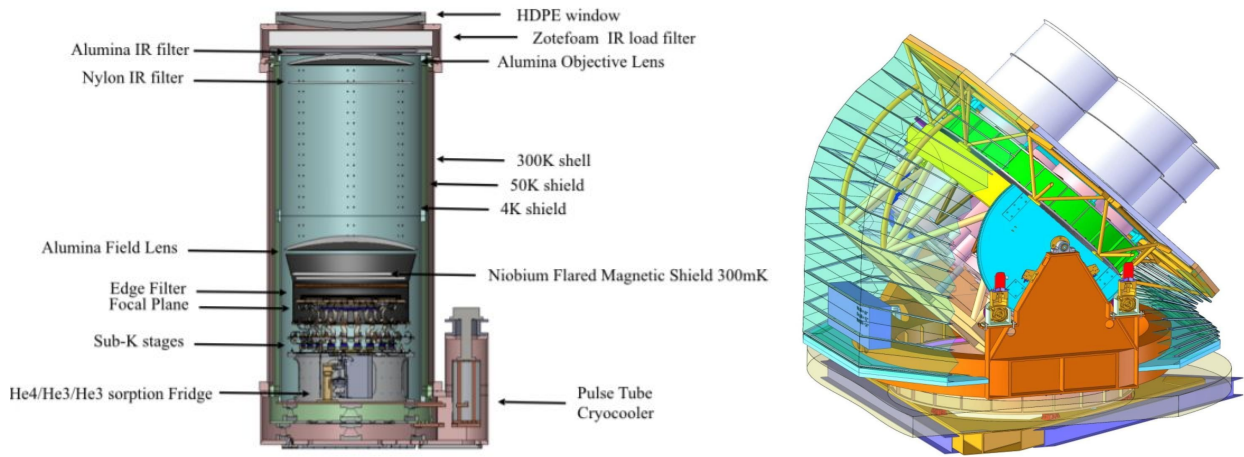

Figure. 1: *Left:* Side view of the Bicep Array receiver which consists of HDPE window, zotefoam IR filter, alumina filter, alumina lenses, nylon IR filter, and sub-K stages. The focal plane houses the detector modules, which are surrounded by several layers of magnetic shielding. *Right:* BICEP Array will use a 3-axis telescope mount capable of rotation in elevation, azimuth, and about the boresight of array.

Each focal plane in BICEP Array will be populated with 12 detector modules. For the first receiver, we adopted a checkerboard layout to cover 30 and 40 GHz with 6 modules each, as shown in Figure 2. The module packages the detector tile and its readout electronics in a compact and magnetically-shielded volume. The module is largely based on the successful BICEP3 design, with the important difference that the detector tile is now fabricated on a 6" silicon wafer, compared to the previous 4", to expedite lithographic fabrication.

Each module contains a 0.625mm thick silicon wafer with an array (5×5 at 40 GHz, 4×4 at 30 GHz) of back-illuminated slot antenna pixels sensitive to vertical and horizontal polarizations. A $\lambda_g/4$ thick quartz tile over the silicon serves as an anti-reflection coating for the incoming radiation. A $\lambda_g/4$ backshort cavity underneath the tile maximizes the radiation coupling to the antennas [8], where $\lambda_g$ is the guided wavelength ($\lambda/n$), $\lambda$ is the free-space wavelength and n is the material index. Each pixel contains two orthogonally polarized detectors A and B, each consisting of a 8×8 slot-antenna array coherently combined through a summing network and whose output is connected to a superconducting transition edge sensor (TES) bolometer. An on-chip band-defining filter with ~25% bandwidth selects the frequency of interest. A superconducting quantum interference device (SQUID) is inductively coupled to the voltage-biased TES sensor and used as an ammeter for the first of two stages of amplification that comprise the multiplexed detector readout. Figure 3 shows a view of the BICEP Array modules at 30 and 40 GHz, which have identical outer dimensions.

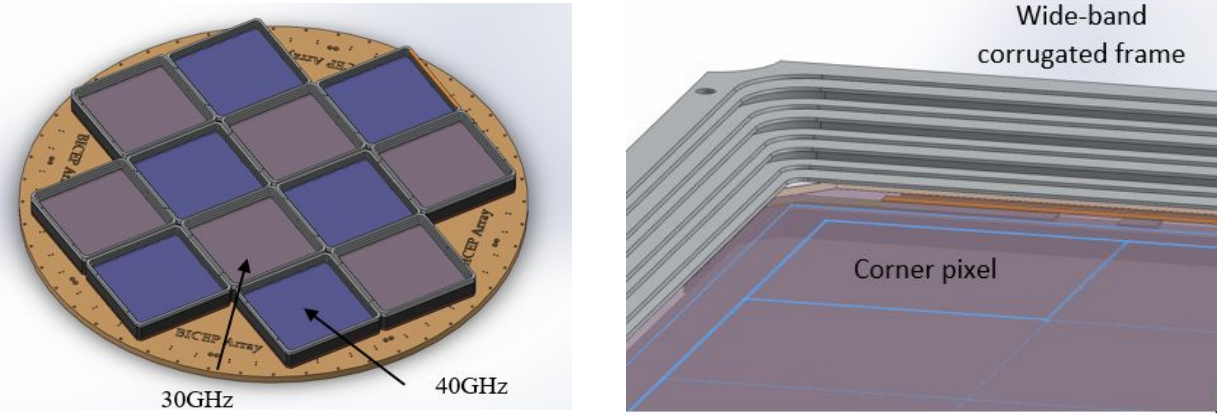

Figure 2: *Left:* View of the 550mm low-frequency BICEP Array focal plane, with the checkerboard layout of 30 and 40 GHz detector modules. *Right:* Wide-band corrugated frame with $3\lambda/8$ distance to the edge pixel.

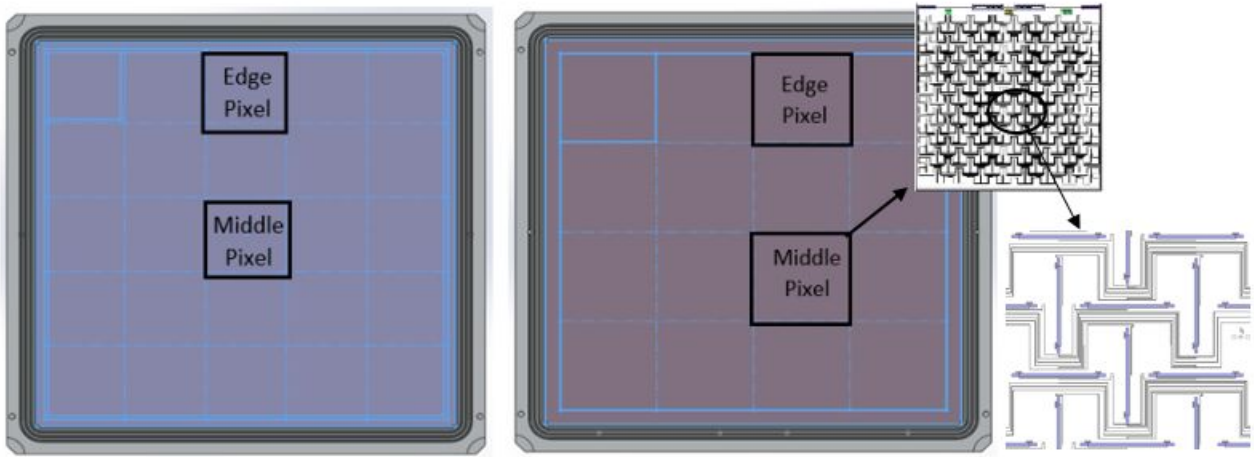

Figure 3: *Left:* Detector module at 40 GHz containing 5×5 pixels. *Right:* Detector module at 30 GHz containing 4×4 pixels. Each pixel has two orthogonal arrays of 8×8 antenna slots for dual polarization observations.

The detector array, anti-reflection wafer, back-short and readout electronics are housed in a Nb box with an aluminum frame around the wafer edges. However, the frame will interact electromagnetically with the pixels located at the edge. The interaction with the frame causes unwanted polarization-dependent beam distortions. We have developed a corrugated frame to reduce beam distortions while maintaining a compact profile to fill the focal plane efficiently. The corrugated frame is designed with carefully optimized groove spacing and depth, as well as vertical and horizontal distance from the antenna to the frame. The BICEP3 design, which was developed prior to this optimization, showed effects of beam steering and temperature to polarization leakage [1]. The authors in [7] defined the differential polarized beam effects as one of the systematic effects which distort the observed B-mode signal. These effects were visible in the beam maps taken in the far field of the detector antenna for edge pixels. The measurements notably show a differential beam effect between two orthogonal polarization detectors, a shift between the A and B beam centers. The difference beam map shows a strong dipole/quadrupole difference beam pattern which leaks temperature to polarization in the CMB polarization maps and produces a false B-mode signal [1-4]. At the data analysis level, we developed a deprojection technique [4] to effectively remove low-order beam effects, but higher-order beam effects could still be present. We aim to minimize these interactions in the BICEP Array receivers, especially in the low-frequency one, since most of its larger-area pixels are located next to the metal frame.

In this paper, we simulate the antenna interactions with a frame. Then, we use the corrugations to minimize the beam mismatches in the far-field of the antenna. We first show the results for a 40 GHz BICEP Array test detector tile coupled to a 25%-bandwidth corrugated frame that was simply scaled from the BICEP3 version at 95GHz. Then, we developed a wide-band corrugation design with a 57% bandwidth that can be used for the 30 GHz and 40 GHz BICEP Array receiver modules, suitable for either a dual-band 30/40 antenna or for single bands, to ease production and assembly. This will help minimizing the far-field pointing mismatch and reduce the T-P leakage in our CMB maps.

## 2. CORRUGATED FRAME DESIGN

When an antenna is in the proximity of a conductive metal frame, interactions between the radiation pattern of the antenna and the conductive frame result in unwanted coupling, which alters the beam pattern of the antenna. These interactions result in highly polarized edge effects, which cause polarized beam mismatch, as shown in Figure 4.

In order to reduce these unwanted interactions, we designed a corrugated surface on the conductive frame with $\lambda/4$ depth and $\lambda/4$ pitch. The grooves change the properties of the conductive wall so as to minimize reflections and beam mismatches. Grooves are usually modeled by treating the corrugations as quarter-wavelength transmission lines [10]. Each corrugation in the frame acts as a short-circuited waveguide with one quarter wavelength depth to a surface wave traveling across it, so that the reflected wave has the opposite phase from the incident wave, thus cancelling the reflected wave. Therefore the corrugated frame acts like free space to waves travelling across it, rather than a short circuit.

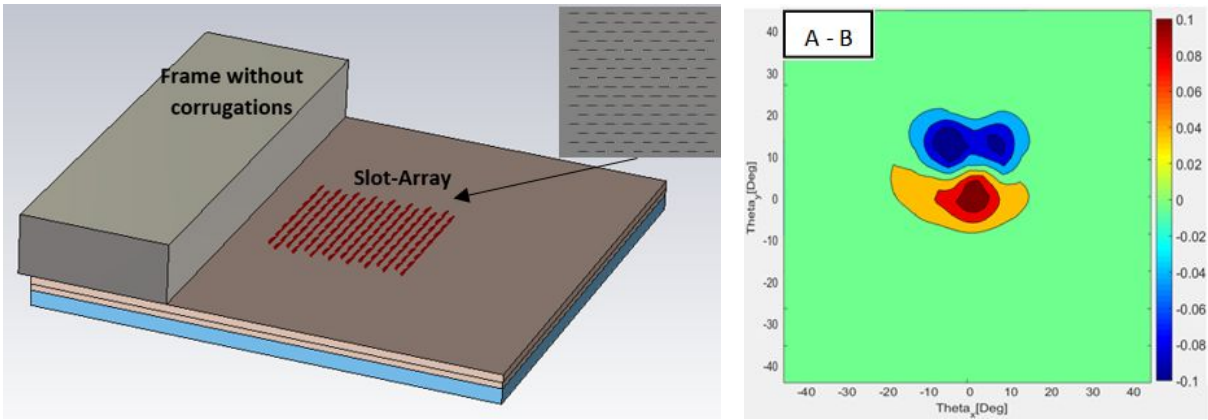

Figure 4: *Left:* Slot-antenna array with all dielectric stacks and with the metal frame (no corrugations) modeled in the CST Microwave Studio. The flat frame is located at $3\lambda/8$ from the slot-antenna array. *Right:* Beam map difference over ~ 25% bandwidth.

### 2.1 Single-Band Corrugated Frame

#### 2.1.1 Simulated and Measured Results

We carried out the numerical analyses using the transient solver in the CST Microwave Studio®, which is based on a finite difference time domain (FDTD) method. Figure 5 shows the simulated antenna-coupled detector with a corrugated frame for one polarization at 40 GHz, used in initial beam simulations. As shown in Figure 5, the corrugations are parameterized by the phase (h), corrugation depth (L), corrugation step (P), corrugated frame angle (Ө) and distance between the antenna and the frame (D). The field patterns are exported to Matlab to plot the contour beams of the vertical polarization (A) and horizontal polarization (B) pixels, as well as the beam difference A–B. We compare these simulations with measurements performed with this geometry. For the initial tests at 40 GHz, we used a modified Keck Array focal plane that was retrofitted with a version of the corrugations that were simply scaled from the BICEP3 version at 95GHz.

The measurement setup is shown in Figure 6. We scan a chopped (10-25 Hz) thermal source in a grid pattern in the far-field of the antennas. This setup has no optics in front of the detectors, except low-reflectivity thermal filters and vacuum window. This allow us to directly measure the far-field pattern of the antenna response without convolving with imaging optics. We measured the beam maps using the detectors on the aluminum superconducting transition [8], designed for the higher optical load during laboratory testing. We demodulate the time-stream data using the chopper reference signal to eliminate noise bias. We repeated the measurement to ensure consistency and high signal-to-noise.

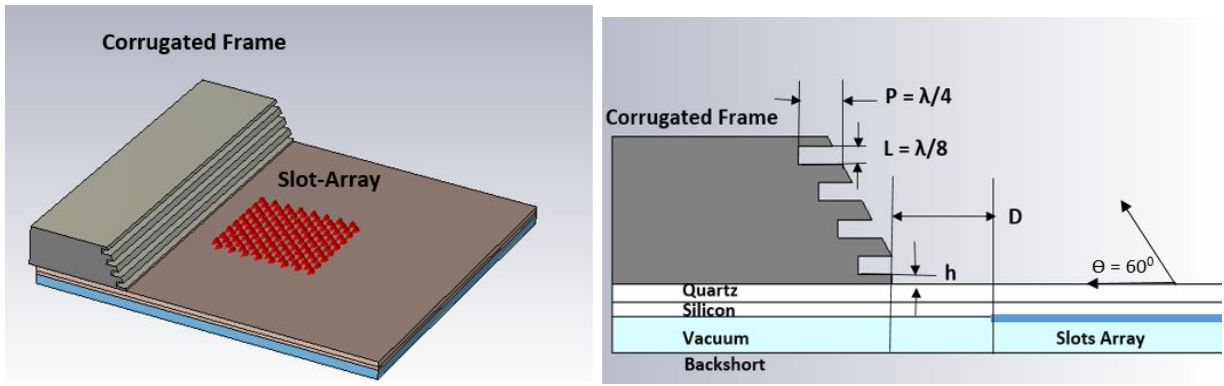

Figure. 5: *Left:* Slot-antenna array with all dielectric stacks and with the corrugated metal frame modeled in the CST Microwave Studio. *Right:* Side view of the simulated model with all optimization parameters. The corrugated frame angle is set to $\Theta = 60°$.

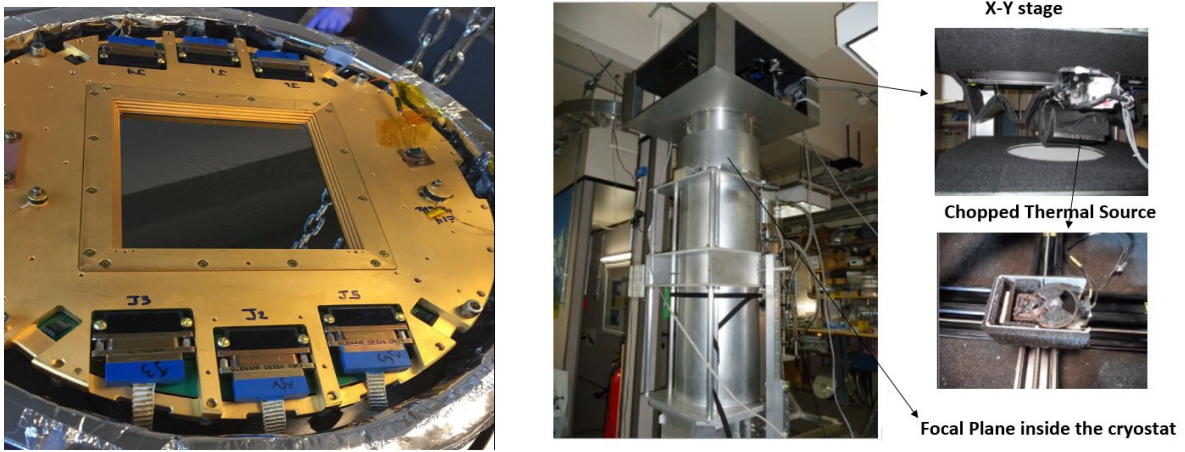

Figure 6: *Left:* Modified Keck Array focal plane used for initial tests at 40 GHz. *Right*: The beam mapper setup, used to measure the angular response of the detectors. The chopper reference cable is connected directly to the Multi-Channel Electronics (MCE) system to enable synchronous detector lock-in demodulation and minimize noise bias.

The simulated and measured beam pattern of a vertically-polarized (A) middle pixel are shown in Figure 7. These beam maps are peak normalized. The simulated and measured beam mismatch are presented in Figures 8 and 9. The simulated and measured beam map difference of the two working middle pixel pairs show a 5% peak-to-peak response in the A-B map. The A-B pattern is approximately quadrupolar, as the A and B beams are both slightly elliptical. The ellipticity is caused by the elliptical beam of single slot, and the finite number of slots (8×8) used in the antenna. This effect could be reduced by using more slots, or choosing X and Y spacings.

Unfortunately, the scaled BICEP3 corrugated frame design used at 95GHz, was not optimal. We simulated the antenna far-field beam map of A-B for this design, reproducing the measured strong 12% peak-to-peak difference map for an edge pixel, as shown in Figure 9, which is consistent with the simulated one.

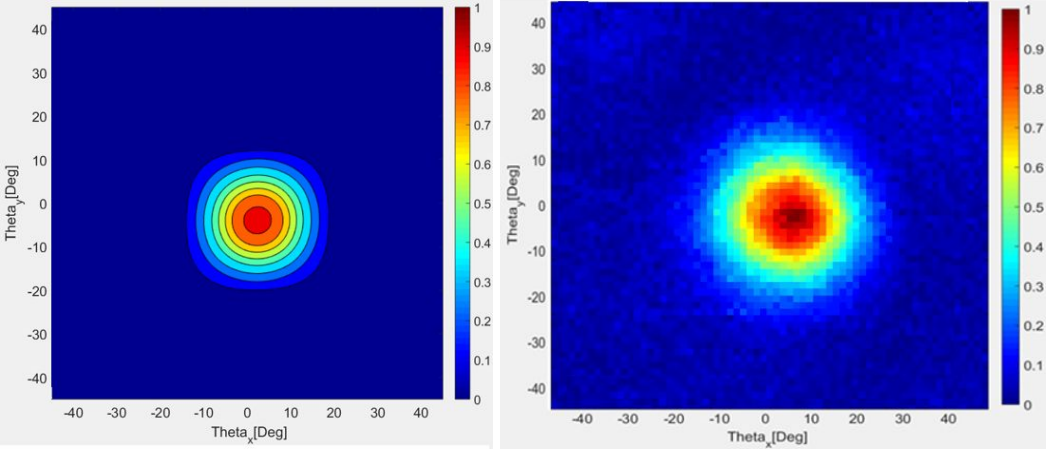

Figure 7: Simulated (*Left*) and measured (*Right*) beam map of vertical polarization (A) for one middle pixel. The beams are simulated in CST and exported to Matlab for contour plot. The beams are peak normalized.

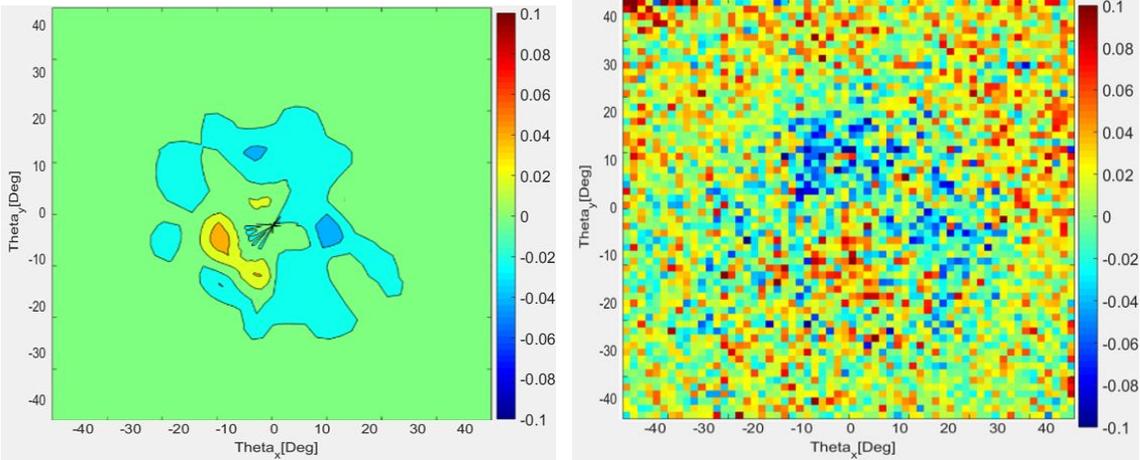

Figure 8: Simulated (*Left*) and measured (*Right*) beam difference map of a working A-B pair for a pixel located in the middle of the detector tile, away from the metal frame, this shows a 5 % peak-to-peak difference.

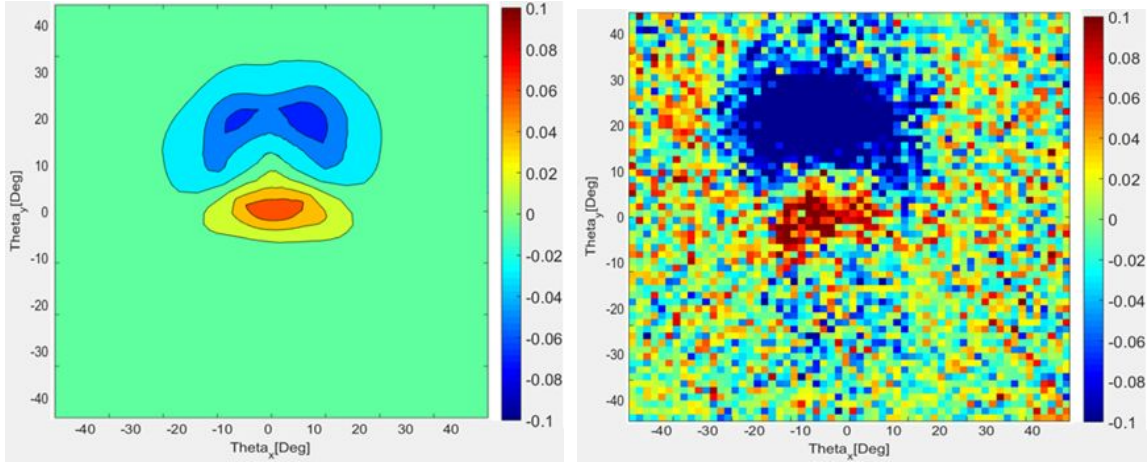

Figure 9: Simulated (*Left*) and measured (*Right*) beam difference map of a working A-B pair for a pixel located at the edge of the detector tile, adjacent to the metal frame, this shows a 12% peak-to-peak difference. The negative difference part of the beam shows a differential gain systematics in addition to the differential pointing.

### 2.1.2 Design Optimization

We optimized the corrugation parameters of the BICEP Array module frame, shown in Figure 5, to minimize the effect of residual beam systematics levels over 25% bandwidth, which is what the detector averages over in nominal science operations. Figure 10 shows the peak-to-peak polarization A-B response averaged over 25% bandwidth as a function of the distance between the antenna and the frame (D), while keeping the other parameters fixed. We repeat the optimization process for all 4 parameters and find h = 0.25 mm, P = $\lambda/4$, L = $\lambda/8$ and D = $3\lambda/8$, where $\lambda$ is the wavelength of the band center of the detector, 7.5 mm. The peak-to-peak difference is about 5% at 40 GHz, over 25% bandwidth, as shown in the beam map in Figure 10 (*right*).

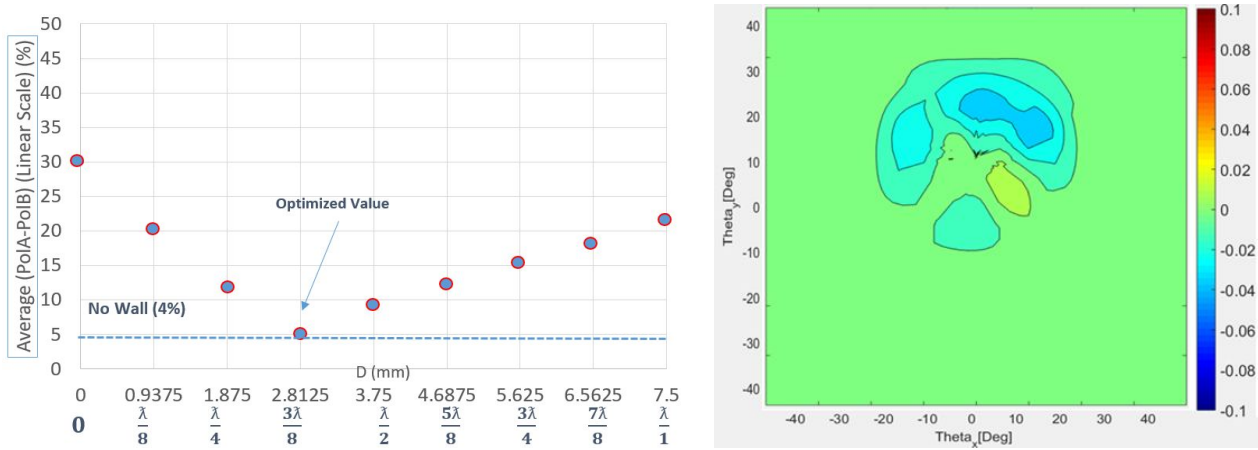

Figure. 10: *Left:* Band-averaged A-B peak-to-peak versus the optimization parameter (D). *Right:* Simulated beam map difference A-B of the optimized design over 25% bandwidth. Similar plots were produced for all 4 corrugation parameters, but are shown here. The dash line indicates the minimum difference due to the intrinsic finite antenna slots in the detector.

The peak-to-peak polarization subtraction versus frequency is shown in Figure 11. The band-averaged A-B peak-to-peak amplitude for an isolated antenna is 4%, and 12% for an edge pixel with a scaled BICEP3 frame. The beam difference of the optimized corrugated frame shows 4% at the single frequency of 40 GHz, similar to that of the isolated antenna, and 5% averaged over 25% bandwidth. The optimization improves the beam mismatch due to the corrugated frame compared to the scaled BICEP3 design.

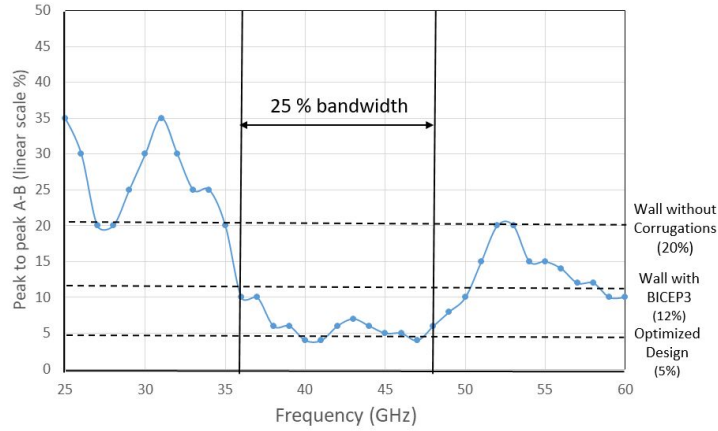

Figure. 11: A-B polarization difference amplitude vs. frequency for the frame without corrugations (20%), frame with scaled BICEP3 corrugations (12%), and the optimized design (5%). For reference, we show the 25% band width that the detectors average over.

## 2.2 Wide-Band Corrugated Frame Design

We also designed a corrugated frame to cover a wider bandwidth by combining two quarter-wavelength corrugated depths. This approach excites two distinct modes and overlapping these modes results in broadband performance. We developed the wide-band corrugated frame for the low-frequency BICEP Array receiver to cover both the 30 and 40 GHz bands, such that a single frame design is suitable for either single frequency. Therefore, the corrugations are ideal for a dual-frequency 30/40 GHz antenna under development. The double-corrugation design is shown in Figure 12.

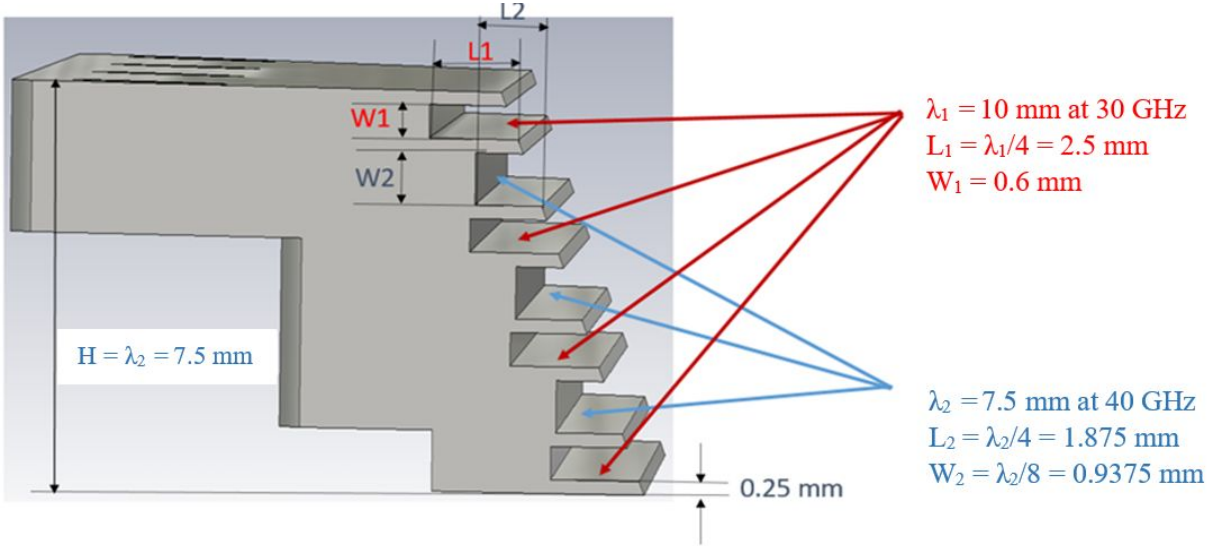

Figure. 12: Side view of wide-band corrugation design in CST Microwave Studio. The design contains four corrugations designed with depth $L_1 = \lambda_1/4$, where $\lambda_1$ is the wavelength at the band center for 30 GHz. The design also contains three corrugations designed with depth $L_2 = \lambda_2/4$, where $\lambda_2$ is the wavelength at the band center for 40 GHz. The corrugation optimized height is $\lambda_2 = 7.5$ mm at 40 GHz.

The peak-to-peak A-B difference versus frequency of the wide-band corrugated frame is shown in Figure 13, covering the BICEP Array bands at 30/40 GHz with a single design. The distance (D) between the antenna and the corrugated frame is optimized to be $3\lambda/8$, where $\lambda$ is the wavelength at the band center of 30 GHz and 40 GHz, respectively. This design minimizes the differential ellipticity to ~ 7% over 57% bandwidth (25 to 45 GHz) for a dual-band 30/40 GHz antenna with D= $3\lambda/8$, where $\lambda$ is the wavelength at the band center of 35 GHz. The wide-band corrugated frame has been fabricated and is currently under testing, as shown in Figure 14.

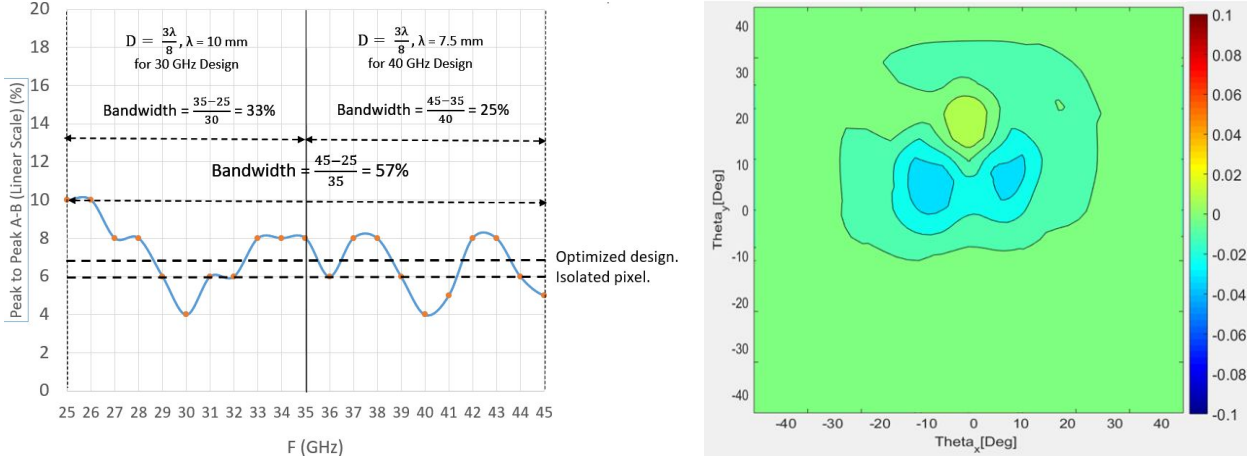

Figure. 13: *Left:* peak-to-peak A-B difference versus frequency of the antenna with the wide-band corrugated frame over 25% and 57% bandwidth. *Right:* The simulated beam map A- B difference of the optimized wide-band design over 57% bandwidth to cover the 30 GHz and 40 GHz designs. The two dash lines indicate the band-averaged difference for the isolated pixel and the optimized design.

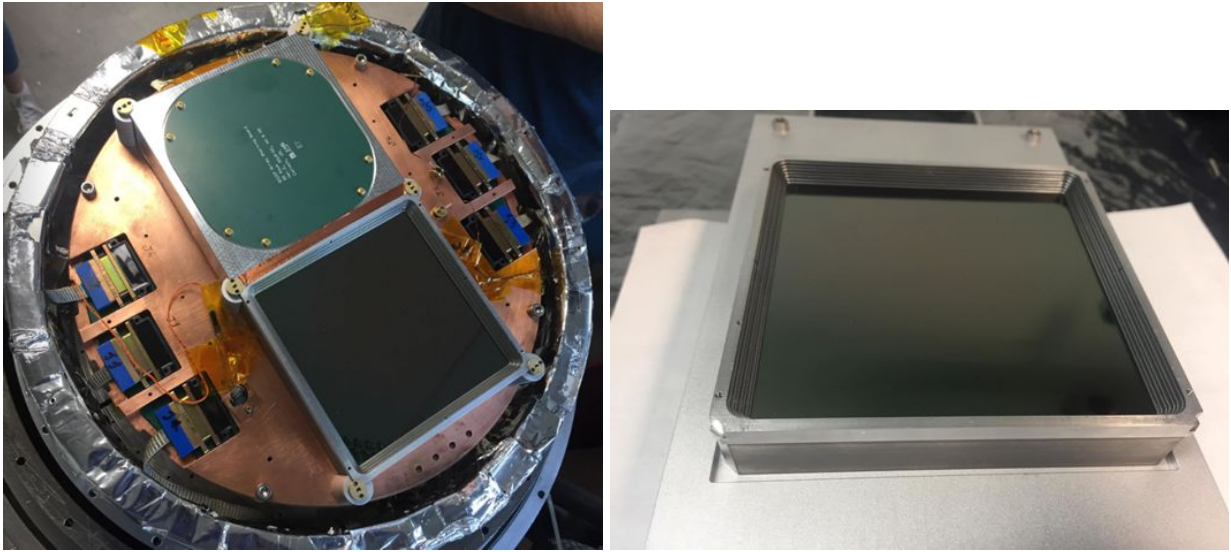

Figure. 14: Fabricated 40 GHz BICEP Array module with wide-band corrugated frame, which is under testing at Caltech. The detector module is cooled down to 280mK by a sorption fridge. We expected to deploy the current design to South Pole at the end of 2019.

## 3. Conclusions

We describe the design and optimization of a 57% bandwidth corrugated frame for the BICEP Array The simulated results show a reduction in residual beam polarized mismatch compared to the previous design used for BICEP3. The wide-band corrugated frame covers 25 GHz to 45 GHz. The initial lab results for the previous BICEP3 show very good agreement with the numerical calculations. The wide-band array module is currently under testing.

## Acknowledgments
The Bicep/Keck Array project have been made possible through a series of grants from the National Science Foundation including 0742818, 0742592, 1044978, 1110087, 1145172, 1145143, 1145248, 1639040, 1638957, 1638978, 1638970, & 1726917 and by the Keck Foundation. The development of antenna-coupled detector technology was supported by the JPL Research and Technology Development Fund and NASA Grants 06-ARPA206- 0040, 10-SAT10-0017, 12-SAT12-0031, 14-SAT14-0009 & 16-SAT16-0002. The development and testing of focal planes were supported by the Gordon and Betty Moore Foundation at Caltech. Readout electronics were supported by a Canada Foundation for Innovation grant to UBC. The computations in this paper were run on the Odyssey cluster supported by the FAS Science Division Research Computing Group at Harvard University. The analysis effort at Stanford and SLAC is partially supported by the U.S. DoE Office of Science. We thank the staff of the U.S. Antarctic Program and in particular the South Pole Station without whose help this research would not have been possible. Tireless administrative support was provided by Kathy Deniston, Sheri Stoll, Irene Coyle, Donna Hernandez, and Dana Volponi. We are grateful our BICEP/Keck Array collaboration colleagues for useful discussions and technical feedback.

## REFERENCES

[1] BICEP2 and Keck Array Collaborations, Ade, P. A. R., Ahmed, Z., Aikin, R. W., Alexander, K. D., Barkats, D., Benton, S. J., Bischoff, C. A., Bock, J. J., Brevik, J. A., Buder, I., Bullock, E., Buza, V., Connors, J., Crill, B. P., Dowell, C. D., Dvorkin, C., Duband, L., Filippini, J. P., Fliescher, S., Golwala, S. R., Halpern, M., Harrison, S., Hasselfield, M., Hildebrandt, S. R., Hilton, G. C., Hristov, V. V., Hui, H., Irwin, K. D., Karkare, K. S., Kaufman, J. P., Keating, B. G., Kefeli, S., Kernasovskiy, S. A., Kovac, J. M., Kuo, C. L., Leitch, E. M., Lueker, M., Mason, P., Megerian, K. G., Netterfield, C. B., Nguyen, H. T., O'Brient, R., Ogburn, IV, R. W., Orlando, A., Pryke, C., Reintsema, C. D., Richter, S., Schwarz, R., Sheehy, C. D., Staniszewski, Z. K., Sudiwala, R. V., Teply, G. P., Thompson, K. L., Tolan, J. E., Turner, A. D., Vieregg, A. G., Weber, A. C., Willmert, J., Wong, C. L., and Yoon, K. W., "BICEP2/Keck Array V: Measurements of B-mode Polarization at Degree Angular Scales and 150 GHz by the Keck Array," Astrophys. J. 811, 126 (Oct. 2015).

[2] BICEP2 Collaboration, Ade, P. A. R., Aikin, R. W., Barkats, D., Benton, S. J., Bischoff, C. A., Bock, J. J., Brevik, J. A., Buder, I., Bullock, E., Dowell, C. D., Duband, L., Filippini, J. P., Fliescher, S., Golwala, S. R., Halpern, M., Hasselfield, M., Hildebrandt, S. R., Hilton, G. C., Hristov, V. V., Irwin, K. D., Karkare, K. S., Kaufman, J. P., Keating, B. G., Kernasovskiy, S. A., Kovac, J. M., Kuo, C. L., Leitch, E. M., Lueker, M., Mason, P., Netterfield, C. B., Nguyen, H. T., O'Brient, R., Ogburn, R. W., Orlando, A., Pryke, C., Reintsema, C. D., Richter, S., Schwarz, R., Sheehy, C. D., Staniszewski, Z. K., Sudiwala, R. V., Teply, G. P., Tolan, J. E., Turner, A. D., Vieregg, A. G., Wong, C. L., Yoon, K. W., and Bicep2 Collaboration, "Detection of B-Mode Polarization at Degree Angular Scales by BICEP2," Physical Review Letters 112, 241101 (June 2014).

[3] BICEP2 and Keck Array Collaborations, Ade, P. A. R., Aikin, R. W., Barkats, D., Benton, S. J., Bischoff, C. A., Bock, J. J., Bradford, K. J., Brevik, J. A., Buder, I., Bullock, E., Dowell, C. D., Duband, L., Filippini, J. P., Fliescher, S., Golwala, S. R., Halpern, M., Hasselfield, M., Hildebrandt, S. R., Hilton, G. C., Hui, H., Irwin, K. D., Kang, J. H., Karkare, K. S., Kaufman, J. P., Keating, B. G., Kefeli, S., Kernasovskiy, S. A., Kovac, J. M., Kuo, C. L., Leitch, E. M., Lueker, M., Megerian, K. G., Netterfield, C. B., Nguyen, H. T., O'Brient, R., Ogburn, IV, R. W., Orlando, A., Pryke, C., Richter, S., Schwarz, R., Sheehy, C. D., Staniszewski, Z. K., Sudiwala, R. V., Teply, G. P.,

Thompson, K., Tolan, J. E., Turner, A. D., Vieregg, A. G., Weber, A. C., Wong, C. L., and Yoon, K. W., "BICEP2/Keck Array. IV. Optical Characterization and Performance of the BICEP2 and Keck Array Experiments," Astrophys. J. 806, 206 (June 2015).

[4] The Keck Array, BICEP2 Collaborations: P. A. R. Ade, Z. Ahmed, R. W. Aikin, K. D. Alexander, D. Barkats, S. J. Benton, C. A. Bischoff, J. J. Bock, R. Bowens-Rubin, J. A. Brevik, I. Buder, E. Bullock, V. Buza, J. Connors, B. P. Crill, L. Duband, C. Dvorkin, J. P. Filippin, S. Fliescher, J. Grayson, M. Halpern, S. Harrison, S. R. Hildebrandt, G. C. Hilton, H. Hui, K. D. Irwin, J. Kang, K. S. Karkare, E. Karpel, J. P. Kaufman, B. G. Keating, S. Kefeli, S. A. Kernasovskiy, J. M. Kovac, C. L. Kuo, E. M. Leitch, M. Lueker, K. G. Megerian, T. Namikawa, C. B. Netterfield, H. T. Nguyen, R. O'Brient, R. W. Ogburn IV, A. Orlando, C. Pryke, S. Richter, R. Schwarz, C. D. Sheehy, Z. K. Staniszewski, B. Steinbach, R. V. Sudiwala, G. P. Teply, K. L. Thompson, J. E. Tolan, C. Tucker, A. D. Turner, A. G. Vieregg, A. C. Weber, D. V. Wiebe, J. Willmert, C. L. Wong, W. L. K. Wu, K. W. Yoon "Measurement of Gravitational Lensing from Large-scale B-mode Polarization" Astrophys. J. (June 2016).

[5] The Keck Array, BICEP collaborations, Hui H. et. al, "BICEP Array: a multi-frequency degree-scale CMB polarimeter" in [Millimeter, Submillimeter, and Far-Infrared Detectors and Instrumentation for Astronomy VIII], International Society for Optics and Photonics (2018).

[6] Crumrine M. for the Bicep/Keck Collaboration "Bicep array cryostat and mount design" in [Millimeter, Submillimeter, and Far-Infrared Detectors and Instrumentation for Astronomy VIII], International Society for Optics and Photonics (2018).

[7] Hu, W., Hedman, M. M., & Zaldarriaga, M. "Benchmark Parameters for CMB Polarization Experiments" Physics Review Letter D (2003).

[8] O'Brient, R., Ade, P. A. R., Ahmed, Z., Aikin, R. W., Amiri, M., Benton, S., Bischoff, C., Bock, J. J., Bonetti, J. A., Brevik, J. A., Burger, B., Davis, G., Day, P., Dowell, C. D., Duband, L., Filippini, J. P., Fliescher, S., Golwala, S. R., Grayson, J., Halpern, M., Hasselfield, M., Hilton, G., Hristov, V. V., Hui, H., Irwin, K., Kernasovskiy, S., Kovac, J. M., Kuo, C. L., Leitch, E., Lueker, M., Megerian, K., Moncelsi, L., Netterfield, C. B., Nguyen, H. T., Ogburn, R. W., Pryke, C. L., Reintsema, C., Ruhl, J. E., Runyan, M. C., Schwarz, R., Sheehy, C. D., Staniszewski, Z., Sudiwala, R., Teply, G., Tolan, J. E., Turner, A. D., Tucker, R. S., Vieregg, A., Wiebe, D. V., Wilson, P., Wong, C. L., Wu, W. L. K., and Yoon, K. W., "Antenna-coupled TES bolometers for the Keck array, Spider, and Polar-1," in [Society of Photo-Optical Instrumentation Engineers (SPIE) Conference Series], Society of Photo-Optical Instrumentation Engineers (SPIE) Conference Series 8452 (Sept. 2012).

[9] Kang, J., Ade, P., Ahmed, Z., Alexander, K. D., Amiri, M., Barkats, D., Benton, S., Bischo, C. A., Bock, J., Boenish, H., et al., "2017 upgrade and performance of bicep3: a 95ghz refracting telescope for degree- scale cmb polarization," in [Millimeter, Submillimeter, and Far-Infrared Detectors and Instrumentation for Astronomy VIII], International Society for Optics and Photonics (2018).

[10] Olver A. D., Clarricoats P. J. B. "Corrugated Horns for Microwave Antennas" Book. (1984).